\documentclass[prd, twocolumn, amsmath, amssymb, nofootinbib, superscriptaddress]{revtex4-2}
\pdfoutput=1

\newcommand{\ket}[1]{\left| #1 \right\rangle}

\usepackage{booktabs}
\usepackage{tabularx}
\usepackage{array}
\newcolumntype{L}[1]{>{\raggedright\arraybackslash}p{#1}}
\usepackage{threeparttable}
\usepackage{hyperref}
\usepackage{orcidlink}

\begin{document}

\title{The Topological Multiverse as the Self-Consistent Extension of Everettian QM to Quantum Gravity}

\email{eshaya2@gmail.com}
\author{Edward J. Shaya\, \orcidlink{0000-0002-3234-8699}}
\affiliation{Department of Astronomy, University of Maryland, College Park, USA}

\begin{abstract}
The ``fine-tuning" of the fundamental constants, from the cosmological constant to the gauge structure of the Standard Model, suggests that our universe inhabits a rare, life-permitting island within a vast landscape of theoretical possibilities. We argue that this landscape arises as a self-consistent extension of Everettian quantum mechanics, once the Wheeler--DeWitt path integral is allowed to sum over admissible topologies rather than being restricted to a single background. Enlarging the DeWitt sum of geometries to include all smooth manifolds supporting causal dynamics promotes the dimensionality, gauge groups, and coupling constants from fixed background inputs to dynamical variables of the sum. 

Spacetime is admitted only where metric elements sustain an extended differentiable Lorentzian geometry. Omnipresent quantum metric fluctuations, combined with curvature-induced deformation, can drive a substantial fraction of elementary regions outside the nondegenerate Lorentzian class. Under the adopted admissibility condition these regions are absent from the physical manifold and constitute quantum topological holes (QTHs). A quantum boundary $\mathcal{B}_Q$ occurs in collapse when the surviving Lorentzian regions lose the connectivity required for an extended differentiable spacetime; in the time-reversed expanding interpretation, it marks the onset of admissible spacetime rather than emergence from a singularity or a pre-geometric foam. At $\mathcal{B}_Q$, the integral generates a coherent superposition of distinct topologies that branch into disjoint U-sectors, each governed by its own effective field theory, with inter-sector transitions dynamically frozen shortly after nucleation.
The topological multiverse is, in this view, not imposed but implicit in the quantum mechanics revealed by the sum over geometries.
\end{abstract}

\maketitle

\section{Introduction}
\label{sec:intro}

The empirical fact that the fundamental constants of nature, such as the fine-structure constant $\alpha$, the cosmological constant $\Lambda$, and the Yukawa couplings, occupy an extremely small region of parameter space compatible with large-scale structure, long-lived stars, and planetary biospheres is known as the ``fine-tuning problem." In the 1970s, Carr and Rees~\cite{CarrRees1979} noted that this situation looks far less surprising if many distinct sets of constants are \emph{physically} realized. In the conclusion of their seminal paper, they discuss the
hypothesis of Wheeler that envisages an infinite ensemble of universes:
\begin{quote}
One would have achieved something if one could show that any cognizable universe had to possess some features in common with our Universe.  Such an ensemble of universes could exist in the same sort of space as the Everett picture invokes.
\end{quote}
This paper is, in essence, a rigorous exploration of the mathematical consequences and physical implications of that statement. If the universal wavefunction contains all such branches, we should simply expect to find ourselves in one of the rare, habitable instantiations of theoretical possibilities.

One leading non-teleological approach invokes a multiverse \cite{Susskind2003,BoussoPolchinski2000,Guth2000,LindeVilenkin1984}. Eternal inflation populates a vast ensemble of causally disconnected regions with different effective low-energy laws through stochastic bubble nucleation; brane-world scenarios realize different physics on different three-branes in a higher-dimensional bulk; and the string landscape supplies the discrete fluxes and moduli that fix the low-energy effective theory in each vacuum. Such mechanisms generate genuine diversity, but that diversity is bounded by the dimensionality and algebraic structure of the parent framework: they explore only the landscape reachable within a set of antecedent meta-laws, which themselves remain unexplained. Whether those meta-laws are arbitrary or uniquely forced by consistency is precisely the question a complete account must answer.

In contrast, the topological/MWI framework reduces the antecedent meta-laws to the minimal consistency conditions of quantization itself. The admissibility conditions, constraint-algebra closure, a normalizable Wheeler--DeWitt kernel, and hyperbolic signature are not a choice of dynamics from which a landscape is then populated; they are the minimal conditions for a sector to constitute a self-consistent universe at all. What remains presupposed is only the quantum formalism, unitarity in the bulk and a sum over histories, not any particular field content, dimensionality, or symmetry group. The laws are thereby contingent rather than necessary: a wide ensemble of mutually consistent possibilities is realized, and we inhabit one that happens to satisfy these conditions, with no deeper principle singling it out and no parent framework left unexplained behind it.

The Many-Worlds Interpretation (MWI) in its most literal form applies to the \emph{entire} universe; there is no external observer and no postulate that superpositions of states collapse to a single state. The universal state $\ket{\Psi}$ always evolves unitarily, and what we call U-sectors are emergent, effectively autonomous components of $\ket{\Psi}$ that arise when the environment becomes entangled with (and thereby recorded in) untracked internal degrees of freedom. In each state of the universe, the environment is internal: coarse variables described in this paper (large-scale geometry and matter fields) become entangled with inaccessible microstructure (short-wavelength field modes, gravitational radiation, and other internal sectors), suppressing interference between alternative quasi-classical histories. All decohered semiclassical histories consistent with the universal dynamics are realized within $\ket{\Psi}$, with branching identified with the progressive loss of interference between these histories under coarse-graining (ignoring further minor splitting).

We treat the ensemble of low-energy laws as \emph{intrinsically quantum}, embedded in the wavefunction that describes our universe. To do so, we build upon the deep connection between the geometry of extra dimensions and the effective laws of physics, a concept recognized since the seminal work of Kaluza and Klein~\cite{Kaluza1921,Klein1926}. It is now well understood in high-energy theory that the symmetries of a compact manifold manifest as gauge forces, and its topology dictates the particle spectrum. This relationship was established in string theory, where the specific choice of a Calabi-Yau manifold determines the observable laws, including the number of particle generations and Yukawa couplings~\cite{Candelas1985, Witten1985} by which particles get their mass.

This paper is organized as follows: Sections \ref{sec:QuantumOrigin} and \ref{sec:enlarge} construct the extended configuration space of the effective theory and its direct-sum Hilbert structure. Sections \ref{sec:Topology} and \ref{sec:Kinematics} detail how the Wheeler--DeWitt path integral topologically generates these disjoint U-sectors, followed by a formalization of the hyperdimensional Feynman--DeWitt sum of geometries in Section \ref{sec:hyperdimensional} and its cosmological implications for a universal crunch in Section \ref{sec:Cyclic}. Section \ref{sec:measures} discusses both the lack of requirements on the measure of universes with full complexity and the future possibilities of providing certain conditional probabilities.

\section{The Quantum Origin of Physical Constants}
\label{sec:QuantumOrigin}

A foundational principle of quantum mechanics is that any physical quantity allowed to vary or to be probabilistically realized must be incorporated into the Hilbert-space description of the theory, either as an operator acting on states or as a label of a superselection sector. In contrast, in the standard formulation of quantum field theory and cosmology, the defining parameters of the Standard Model, gauge couplings, masses, mixing angles, and symmetry structure, are not represented within the quantum state at all, but instead enter as fixed numerical inputs in the Lagrangian. This asymmetry points to a conceptual incompleteness.

To resolve this, we elevate these parameters to dynamical quantum variables. In this framework, the ``constants" we observe are not fundamental inputs but derivative quantities determined by the vacuum state of the high-energy theory. While phase transitions such as QCD confinement or electroweak symmetry breaking occur at relatively low energies (post-inflation), the parameters governing these transitions (coupling strengths, potential shapes, and mass scales) are determined by the stabilization of fields at the GUT and Planck scales~\cite{Vilenkin1985} but are predetermined once the manifold is chosen. Table~\ref{tab:constants_classes} summarizes the hierarchy of these parameters.

\begin{widetext}
\begin{table*}[!ht]
\centering
\begin{threeparttable}
\begin{tabular}{L{2.0cm} L{3.5cm} L{11.8cm}}
\hline
\emph{Class} & \emph{Constants Fixed} & \emph{GUT/Planck Mechanism} \\
\hline
\emph{Spacetime \newline Topology} & $D, d, n, c, h, k, \sigma$ & \emph{Quantized Dimensionality:} The total dimension $D$ is fixed by quantum consistency (e.g., anomalies). The macroscopic $d$ are fixed by topological compactification. \\
\hline
\emph{Unified \newline Couplings} & $G, \alpha, \alpha_W, \alpha_s, e, \epsilon_0, \mu_0$ & \emph{Moduli Stabilization:} Scalar fields (Dilaton/Volume moduli) settle into potential minima, fixing the strength of gravity and the unified gauge forces. \\
\hline
\emph{Mass \newline Generation} & $m_e, m_p, R_\infty, M_W$ & \emph{Blueprint Fixing:} The high-energy moduli fix the Yukawa couplings and Higgs potential parameters. This predetermines the masses ($m \sim y \cdot v$) and confinement scales ($\Lambda_{QCD}$) that emerge at lower energies. \\
\hline
\emph{Vacuum \newline Energy} & $\Lambda$ (Cosmological Constant) & \emph {The vacuum energy density.} \\
\hline
\end{tabular}
\caption{Classification of Physical Constants by GUT-Scale Origin. All fundamental parameters are determined by geometric and topological transitions occurring before the onset of cosmic inflation, ensuring their universality across each U-sector.}
\label{tab:constants_classes}
\end{threeparttable}
\end{table*}
\end{widetext}

We treat $c$, $\hbar$, and $k_B$ as the fixed background units of the theory space.
These are not dynamical couplings but unit-conversion factors.
They carry no dimensionless content, and only dimensionless ratios (such as $\alpha$ or mass ratios) constitute the physical variation across sectors. Normalizing $c=\hbar=k_B=1$ establishes the conversion factors between space, time, energy, and information that hold across all sectors.
The remaining dimensionful scale, the Planck scale itself, is set not by these conventions but by gravity, and Newton's constant $G$ therefore requires a distinct treatment.
We take the fundamental $D$-dimensional gravitational constant
$G_D$ to define the common Planck scale. The effective four-dimensional coupling
$G_{\rm eff}$, by contrast, is not fundamental but a derived geometric quantity, fixed within each sector by the volume of the compactified dimensions
$V_{\rm compact}$ (and, in a string realization, the coupling $g_s$).

\subsection{Motivating a Diversity of Algebras}

We pause to make precise the notion of \emph{complexity} that the
anthropic argument will rely on, since it does substantial work in what
follows. We use the term in a layered, operational sense, descending a
hierarchy of capabilities, each of which is a necessary condition for the
next:
\begin{enumerate}

    \item \emph{Structural complexity:} the existence of long-lived
    bound states across a wide range of scales: atomic nuclei,
    atoms, molecules, stars, planetary systems, galaxies. This requires
    a gravitational sector that clumps, an expansion history that is slow
    enough to permit clumping, a confining strong interaction that
    binds nucleons, and a long-range force that admits stable
    multi-scale equilibria.
    
    \item \emph{Chemical complexity:} a sufficiently rich periodic
    table to support combinatorial bonding: many distinct stable
    elements, a covalent regime, and liquid-phase solvents in which macromolecular chemistry can occur. This narrows the allowed ranges of $\alpha$, the electron-to-proton mass ratio, and the
    light quark masses well beyond what mere structural complexity
    requires.
    
    \item \emph{Informational complexity:} the capacity of the
    chemical substrate to support self-replicating, error-correcting,
    memory-bearing systems (the substrate-independent threshold for
    observers in the Everettian sense, namely, entangled subsystems
    capable of recording coarse-grained outcomes and thereby
    participating in decoherent branching).

\end{enumerate}

When the present paper speaks of ``complexity'' without further qualification, we mean the full chain: a U-sector counts as complexity-permitting if and only if its physical laws admit all three tiers.

The fundamental character of physical law is dictated almost entirely by the algebraic structure of symmetry groups. The organization of matter into particle families, chiral structures, and interaction dynamics is a strictly algebraic reflection of the underlying gauge group $\mathcal{G}$. This algebraic dependence imposes a severe constraint on habitability. Complexity requires a delicate balance, such as a confining non-Abelian force ($SU(3)$) to bind nucleons and a long-range Abelian force ($U(1)$) for chemistry. Therefore, a habitable multiverse must sample from a broad landscape of symmetries, finding those rare branches where the algebra breaks into a specific product group capable of supporting information-dense structures.

\section{Enlarged Configuration Space}
\label{sec:enlarge}

\subsection{Superspace and Theory Space}

To formulate a wavefunction of the universe, we must define the domain over which it evolves. We can define it either by its final states or by its initial geometry. First, we look at the final states. Standard quantum cosmology utilizes Wheeler's superspace, $S = \{h_{ij}, \phi\}$, describing the 3-geometry and matter fields.
The geometric structure on this superspace is the DeWitt supermetric
$G^{ijkl}$, whose mathematical form we adopt unchanged from DeWitt's original
construction~\cite{DeWitt1967}. Two structural features of $G^{ijkl}$ matter
for what follows. Its signature is hyperbolic, with the negative direction
corresponding to overall spatial scale and the positive directions to shape
deformations; this is what is meant by ``hyperbolic signature'' in the
admissibility conditions of Sec.~\ref{sec:intro}. And it supplies the natural
measure with respect to which Wheeler--DeWitt kernel normalizability is
defined.

The present framework leaves the formula for $G^{ijkl}$ untouched but applies
it sector by sector. Each sector carries
its own superspace, whose dimensionality is set by the metric degrees of
freedom of the manifold and which in general differs from sector to sector.
Accordingly, the global supermetric on the Grand Hilbert space is
block-diagonal in the sector basis.

However, in a principled quantum context, ``matter" fields are not fundamental inputs but derived consequences of the internal geometry.
Therefore, we redefine the configuration space to eliminate this redundancy. The fundamental degrees of freedom are partitioned into macroscopic geometry and microscopic topology. The \textbf{extended superspace} is defined as the union of the geometric superspace and the theory space:
$\widehat{\mathcal{Q}} \;\equiv\; \text{Riem}(\Sigma) \times \mathcal{T}$.

A point in this space is defined by the coordinate vector:
\begin{equation}
	\widehat{\mathcal{Q}} = \{ h_{ij}(x), \, \mathfrak{U}, \, \mathfrak{D}, \, \mathfrak{X} \}.
\end{equation}

\begin{itemize}
	\item $h_{ij}(x)$ is the observed metric, describing the classical curvature of the macroscopic spatial slice (e.g., an FRW universe with localized perturbations).
	\item $\mathfrak{U}$ denotes the universality class. In the final state, this is the specific set of gauge symmetries and matter representations available (e.g., $SU(3) \times SU(2) \times U(1)$ with three generations). It is the qualitative identity of the physical laws.
	\item $\mathfrak{D}$ denotes the \textbf{dimensional factorization} state. It specifies the partition into macroscopic ($d$) and compact ($n$) dimensions, selected dynamically.
	\item $\mathfrak{X}$ represents the vector of physical constants. These are the scalar values: masses, coupling constants, and the cosmological constant, determined by the specific vacuum expectation values (valleys) at which the moduli stabilized. In this superspace, $\mathfrak{X}$ distinguishes between universes with identical symmetries ($\mathfrak{U}$) but different interaction strengths (e.g., a weak gravity universe vs. a strong gravity universe).
\end{itemize}

\subsection{U--Sector Hilbert Spaces and the Direct-Sum Structure}

We define a \emph{U-sector} not merely as a distinct quantum state, but as a distinct superselection sector within the universal Hilbert space. It is insufficient to treat universes with different physical constants simply as orthogonal vectors within a single, shared Hilbert space. Because the fundamental constants (such as the electron mass or fine-structure constant) appear in the definition of the canonical commutation relations and the Hamiltonian operator itself, states associated with different constants belong to inequivalent representations of the field algebra.

The appropriate mathematical structure, for a purely discrete family of background theories, is a \emph{direct sum} of the Hilbert spaces associated with every possible choice of physical laws~\cite{Roberts1990}. Let $\mathcal{H}^{(\mathfrak{U},\mathfrak{D},
\mathfrak{X})}$ denote the standard Hilbert space of canonical quantum gravity for a fixed background theory $\mathcal{T}=(\mathfrak{U},\mathfrak{D},\mathfrak{X})$. For a discrete theory space $\mathcal{T}$ the total, or ``Grand," Hilbert space is
\begin{equation}
\mathcal{H}_{\rm grand}
    = \bigoplus_{(\mathfrak{U},\mathfrak{D},\mathfrak{X}) \in \mathcal{T}} \mathcal{H}^{(\mathfrak{U},\mathfrak{D},\mathfrak{X})}.
\end{equation}
More generally, when the theory space $\mathcal{T}$ contains continuous families of backgrounds (e.g.\ when some components of $\mathfrak{X}$ are continuous couplings or moduli), the direct sum is replaced by its continuous analog, a direct integral (also known as the Hilbert integral) of Hilbert spaces,
\begin{equation}
\mathcal{H}_{\rm grand}
    = \int_{\mathcal{T}}^{\oplus} \mathcal{H}^{(\mathfrak{U},\mathfrak{D},\mathfrak{X})}\, d\mu(\mathfrak{U},\mathfrak{D},\mathfrak{X}),
\end{equation}
for some natural measure $d\mu(\mathfrak{U},\mathfrak{D},\mathfrak{X})$ on theory space. In this sense a U-sector is always defined by a definite pair $(\mathfrak{U},\mathfrak{D},\mathfrak{X})$, but the set of U-sectors itself may form a continuum.

In this framework, the meta-wavefunctional of the universe is a vector with components in each sector:

\begin{equation}
\Psi = \bigoplus_i \Psi_i, \nonumber\;\;
\Psi_i \equiv \Psi_{\mathfrak{U}_i,\mathfrak{D}_i,\mathfrak{X}_i}, \nonumber\;\;
\Psi_i \in \mathcal{H}^{(\mathfrak{U}_i,\mathfrak{D}_i,\mathfrak{X}_i)}.
\label{eq:Psidecomp}
\end{equation}

We emphasize that the absence of interference between U-sectors is not attributed to ordinary late-time laboratory decoherence alone. Unlike a standard superposition, the components $\Psi_{\mathfrak{U}_i,\mathfrak{D}_i,\mathfrak{X}_i}$ evolve under different Hamiltonians $\mathcal{H}_{\mathcal{T}, \mathfrak{D}_i,\mathfrak{X}_i}$. Consequently, there are no local physical operators that can map states from one sector to another, rendering them physically disjoint (superselected). Operationally, one may simply regard each sector as carrying its own canonical Hilbert space and treat the sectors as mutually isolated theories for all practical purposes.

\section{The Topological Basis of Theory Space}
\label{sec:Topology}

Up to this point, we have treated the Theory Space coordinates $\mathcal{T}(\mathfrak{U},\mathfrak{D},\mathfrak{X})$ as independent variables governing the effective laws. This \emph{effective description} is useful to define the fine-tuning problem in terms of observable quantities ($G_{\rm eff}, \alpha, \Lambda$) and allows us to formulate the probabilistic evolution of the universe in terms of observables.

However, this description is physically incomplete. In the effective basis, the potential $V(\mathfrak{U},\mathfrak{D},\mathfrak{X})$ acts as an external input, a phenomenological ``black box." To identify the physical source of this potential, and thus the mechanical origin of the constants, we must descend to the \emph{generative basis} of theory space $\mathcal{T}(\mathcal{M}, \vec{N})$.

When we refer to the ensemble of manifolds $\mathcal{M}_n$, we refer strictly to the internal spacetime: the compact, spatial dimensions that physically exist at every point of the macroscopic universe. Formally, we regard the total spacetime as a product structure $\mathcal{M} = \mathbb{R}^{d,1} \times \mathcal{M}_n$, where $D = d + n$. While the macroscopic factor $\mathbb{R}^{d,1}$ describes the extended, observable universe, the curled manifold $\mathcal{M}_n$ (e.g., a Calabi-Yau shape) determines the particle spectrum, gauge symmetries, and vacuum energy.

\subsection{From Parameters to Manifolds}

In this fundamental picture, the observable parameters $(\mathfrak{U},\mathfrak{D},\mathfrak{X})$ are not primary degrees of freedom but are derived observables of the underlying topological manifold $\mathcal{M}$ and its flux configuration $\vec{N}$.
By a \emph{cycle} of $\mathcal{M}_n$ we mean a closed, non-contractible
submanifold representing a non-trivial homology class; the cycles of
distinct dimension form the homology of $\mathcal{M}_n$ and are the
topological objects on which the framework's quantization conditions are
imposed. Each cycle carries a quantized flux, defined by the integral of
the relevant field strength over the cycle (Dirac quantization), and a
geometric volume $V_{\rm cyc}$ set by the moduli; the resulting low-energy
couplings, masses, and stabilization energies depend on both the discrete
flux integers $\vec{N}$ and the continuous volumes $\{V_{\rm cyc}\}$.

As established in the foundational work on Calabi-Yau manifolds~\cite{Candelas1985, Witten1985}:
\begin{itemize}
    \item The Gauge Group $\mathfrak{U}$ is dictated by the isometry group and intersection numbers of $\mathcal{M}$.
    \item The Coupling Constants $g_i$ (contained in $\mathfrak{X}$) are determined by the stabilized volumes of specific cycles within $\mathcal{M}_n$.
    \item The internal dimensionality $n$ is a consequence of flux stabilization preventing specific dimensions from decompactifying.
\end{itemize}
The transition from $\mathcal{T}(\mathfrak{U},\mathfrak{D},\mathfrak{X})$ to $\mathcal{T}(\mathcal{M}, \vec{N})$ represents the UV-completion of our framework. The fine-tuning of continuous parameters observed in the effective theory is revealed, at the fundamental level, to be the discrete selection of a stable topological isomer.

\subsection{Geometry Dictates Interaction}

Correspondences between manifolds and gauge groups are well established. As recognized in Kaluza-Klein theory and generalized in string compactifications, the isometries and topological invariants of the hidden spatial dimensions appear to a macroscopic observer as fundamental forces and particle spectra. For instance, the Euler characteristic of a Calabi-Yau manifold dictates the number of particle generations, while its intersection numbers determine Yukawa couplings \cite{Candelas1985, Witten1985}. In this framework, the wavefunction of the universe $\Psi$ populates a vast landscape of these topological sectors. The initial meta-state is a superposition $\Psi_{\text{initial}} = \sum_{\mathcal{M}} c_{\mathcal{M}} | \mathcal{M} \rangle$ over the infinite set of consistent topological spaces, where the macroscopic topology $\mathcal{M}$ is a random variable distributed across the ensemble.

\section{The Manifold Creates U-sectors}
\label{sec:Kinematics}

\subsection{Native Emergence at the \texorpdfstring{$\mathcal{B}_Q$}{BQ} Boundary}
\label{subsec:NativeEmergence}

In the quantum-boundary framework, spacetime exists only where admissible metric elements sustain a connected, differentiable Lorentzian geometry. Omnipresent quantum metric fluctuations, combined with curvature-induced deformation, including tidal shear in anisotropic collapse, can drive individual metric quanta outside the nondegenerate Lorentzian class. We adopt the Lorentzian admissibility condition of the companion analysis: such realizations are excluded from the physical manifold, leaving quantum-sized four-dimensional excisions called quantum topological holes (QTHs) \cite{Shaya2026_Sig,Shaya2026_BQ}. As these holes proliferate, the surviving Lorentzian regions eventually cease to support an extended differentiable spacetime. This loss of support defines the quantum boundary $\mathcal{B}_Q$. In collapse it terminates the spacetime continuation; in the time-reversed nucleation interpretation it marks the onset of an expanding Lorentzian geometry. Different paths by which Lorentzian elements percolate into a connected universe may generate distinct topologies and thereby different effective physical laws.

Homogeneous cosmological collapse provides a counterpart to the tidal mechanism operating in anisotropic geometries. Although an exactly Friedmann geometry has no shear or Weyl curvature, its trace-free Ricci tensor, $S_{\mu\nu}\equiv R_{\mu\nu}-\tfrac14 Rg_{\mu\nu}$, need not vanish. This curvature produces a finite-scale deformation of the metric across a spatial quantum of size $q$. For spatially flat radiation-dominated geometry, we model this response phenomenologically as $s^{(\rm FRW)}_{\hat\mu\hat\nu}=-C_{\rm FRW}\mathcal{T}\,\operatorname{diag}(3,1,1,1)$, where $\mathcal{T}=H^2q^2/c^2$ and $C_{\rm FRW}\simeq1$ is an assumed order-unity response coefficient, rather than a coefficient derived from a normal-coordinate expansion. The illustrative boundary densities are conditional on this response prescription and normalization. Combined with quantum metric fluctuations, the deformation can drive metric elements outside the admissible Lorentzian class, creating quantum topological holes. The cosmological quantum boundary is reached when the remaining Lorentzian elements no longer sustain an extended differentiable spacetime.

For $q=\alpha\ell_P$, the flat Friedmann equation gives $\rho/\rho_P=3\mathcal{T}/(8\pi\alpha^2)$. With the phenomenological normalization $C_{\rm FRW}=1$, homogeneous radiation scans on a periodic box of side $32q$, with one lattice site per $q$ and 16 realizations at each elementary size, give mean individual extinction thresholds $\mathcal{T}_{\rm qb}=0.5994$, $0.6896$, and $0.8026$ for $\alpha=1.9$, $2.5$, and $4$, respectively. The corresponding mean boundary densities are $\rho_{\rm qb}/\rho_P=0.01982$, $0.01317$, and $0.005988$, approximately $2.0\%$, $1.3\%$, and $0.60\%$ of Planck density. Every realization reaches extinction, defined by loss of the last periodically winding component of complete Lorentzian D-state patches, rather than by a 50-percent ensemble survival criterion. These are illustrative results for the adopted finite-scale response; changing $C_{\rm FRW}$ rescales the thresholds in $\mathcal{T}$ and density inversely. Since this response depends on $H^2$, the local signature criterion is the same for contraction and the time-reversed expanding branch \cite{Shaya2026_Sig}.

\subsection{Lorentzian Support and the Quantum Boundary}
\label{subsec:WDWDomain}

The dynamical object in the gravitational sum over histories is the geometry, including its metric, causal structure, and connectivity. We adopt the admissibility condition that nondegenerate Lorentzian metric elements belong to the physical manifold, whereas non-Lorentzian or degenerate realizations are excluded and represented by quantum topological holes (QTHs) \cite{Shaya2026_Sig}. These holes are absent elementary regions, rather than matter fields placed within an otherwise intact background.

Quantum fluctuations are present throughout the geometry. A curvature-induced finite-scale deformation changes the probability that an elementary metric remains Lorentzian. As a substantial fraction of elementary regions becomes inadmissible, QTHs proliferate and the surviving Lorentzian elements cease to sustain an extended differentiable continuum. Individual admissible islands can remain; the boundary does not require every metric element to fail. In the numerical model, the defining event is loss of the last periodically winding component of complete Lorentzian D-state patches. Neither a prescribed one-point Lorentzian fraction nor a 50-percent survival probability defines this event.

Local derivatives and field equations can remain meaningful on surviving differentiable islands, but those islands no longer furnish a continuing macroscopic spacetime. The quantum boundary $\mathcal{B}_Q$ is the boundary of the class of histories supporting that extended differentiable Lorentzian continuation. It is therefore a signature-and-connectivity boundary. The action need not fall to $\hbar$, and loss of a stationary-phase saddle is not its defining mechanism.

Within an admissible semiclassical branch, stationary-phase and WKB methods retain their usual role: the phase of the wavefunctional describes an approximately classical geometry and supports an emergent time, while a Born--Oppenheimer treatment recovers approximate Schr\"odinger evolution for internal degrees of freedom. In the cosmological description used here, this large-scale semiclassical branch is taken to persist up to the signature-connectivity boundary. Saddle methods can describe semiclassical evolution within the admissible domain; the physical restriction at $\mathcal{B}_Q$ follows from loss of differentiable Lorentzian support.

For contraction, $\mathcal{B}_Q$ terminates the spacetime continuation once the surviving support loses extended connectivity. In the time-reversed nucleation interpretation, it marks the onset of an expanding connected Lorentzian geometry. The sum over admissible histories can contain different topologies and effective dimensionalities, producing the distinct U-sectors discussed here. No continuous physical spacetime on the inadmissible side is assumed to precede nucleation or follow terminal collapse. The block universe of each sector begins and ends where an extended differentiable Lorentzian continuation first and last exists (Sec.~\ref{sec:Cyclic}).

\subsection{The Action and Its Sector-Dependent Constraint}
\label{subsec:SectorAction}

Once this smooth manifold is established at the $\mathcal{B}_Q$ boundary, the
path integral acts as a stringent consistency filter. The universal state is a
wavefunctional over possible geometric and internal configurations, weighted by
the phase $e^{iS/\hbar}$. Within the admissible class of geometries, locality and diffeomorphism invariance severely restrict the geometric component of the action. 
Diffeomorphism invariance is not an external stipulation but the requirement
that the geometry's dynamics be independent of the coordinate frame used to
describe it; a geometry whose induced action depended on that arbitrary choice
would be self-inconsistent. Diffeomorphism invariance alone does not single
out a unique action: curvature-squared invariants, Lovelock terms in $D > 4$,
and general $f(R)$ functionals are all frame-independent. It does, however,
organize the admissible terms into a derivative expansion. Control of this expansion requires curvature and gradient corrections to remain small on the surviving smooth regions. The quantum boundary is determined by loss of extended Lorentzian connectivity under quantum fluctuations and curvature-induced deformation; it does not impose a universal Planck-curvature cap or by itself guarantee control of the derivative expansion.
At leading order in this expansion, the geometric action
$S_{\mathrm{geom}}$ is the $D$-dimensional
Einstein--Hilbert action. Because the path integral is evaluated from a hard
topological boundary ($\mathcal{B}_Q$), this action natively requires the
Gibbons--Hawking--York boundary term to ensure a well-posed variational
principle:

\begin{equation}
\begin{split}
S_{\mathrm{geom}} = \frac{1}{16\pi G_D} \int_{\mathcal{M}} R \sqrt{-g}\, d^Dx \\
+ \frac{1}{8\pi G_D} \int_{\mathcal{B}_Q} K \sqrt{|h|}\, d^{D-1}x
\end{split}
\end{equation}

where $G_D$ is the fundamental gravitational constant in $D$ dimensions, $h$
is the determinant of the induced metric on the boundary, and $K$ is the trace
of the extrinsic curvature. Thus, this total geometric action is automatically
selected to weight the macroscopic geometric measure $\mathcal{D}(g_D)$.

The effective action governing the matter and flux measure $\mathcal{D}(\phi)$, conversely, is not uniquely constrained by diffeomorphism invariance alone.
Crucially, this is not the standard perturbative measure of local quantum
field theory evaluating fluctuations around a fixed vacuum; rather, it is a full, non-perturbative measure over all discrete topological and flux sectors.
It is not externally prescribed; it emerges strictly from the allowed topological variants and quantized fluxes evaluated by the integral over the boundary.

There is no single, monolithic Hamiltonian governing the universal state.
Because $\Psi$ contains the superposition of all possible topological
manifolds, the specific form of the Hamiltonian (its dimensionality, its
symmetry groups, and its constraint algebra) is fundamentally determined by
the topological sector itself. Each distinct manifold $\mathcal{M}$ dictates
its own specific Hamiltonian operator $\hat{H}_{\mathcal{M}}$. The
Wheeler--DeWitt constraint is therefore not a single global equation, but a
family of sector-specific constraints
($\hat{H}_{\mathcal{M}} \Psi_{\mathcal{M}} = 0$).

A clarifying remark is in order. The multiplicity of Hamiltonians
$\{\hat{H}_{\mathcal{M}}\}$ does not violate the standard quantum-mechanical postulate of a unique generator of evolution. There is still only one global
operator. It is the total constraint $\hat{\mathcal{H}}$ acting on the Grand
Hilbert space
$\mathcal{H}_G = \bigoplus_{\mathcal{M},\vec{N}} \mathcal{H}_{\mathcal{M},\vec{N}}$.
What we denote $\hat{H}_{\mathcal{M}}$ is simply its restriction to the
sector $\mathcal{H}_{\mathcal{M},\vec{N}}$. The restriction is well-defined because topological labels are superselected: no physical observable has
matrix elements between sectors of distinct $(\mathcal{M},\vec{N})$. The
global operator is therefore block-diagonal in the sector basis. This is the
canonical-gravity analog of an ordinary Hamiltonian decomposed by a
conserved charge. One operator; many blocks; one block per conserved label.

This sector-wise structure is, in fact, standard at the level of canonical
gravity. In a Dirac-quantized generally covariant theory, the Hamiltonian is
itself a constraint, $\hat{H}\Psi = 0$, and the form of that constraint
depends on the topological data of the configuration on which the fields
are defined. The new content of the present framework lies elsewhere. It is the interpretation of the direct-sum wavefunction
$\Psi = \sum c_{\mathcal{M},\vec{N}} |\mathcal{M},\vec{N}\rangle$ as a
physically real superposition over admissible sectors, rather than a formal device for labeling separate theories.

This sector-dependent filter ensures that only those components where the
topological structure successfully balances the gravitational and matter/flux
contributions can survive to form a viable, expanding semiclassical branch.
Inflation, in some sectors, then exponentially enlarges this one domain, homogenizing that same
$(\mathcal{M},\vec{N})$ across an enormous physical volume. Regions beyond our
present horizon are therefore not expected to realize different laws; they
are simply causally inaccessible portions of the same branch.

\subsection{Suppression of Inter-Sector Transitions}
\label{subsec:Freezeout}

A natural worry about this picture is whether the individual manifolds have
sufficient time,  after $\mathcal{B}_Q$, to transition between alternative topologies and reach an equilibrium distribution across the landscape. Such transitions would be destructive as well as redistributive: a sector that had settled into a rare, complexity-permitting configuration could tunnel away from it, erasing the very structure the framework requires for observers.
We show that this does not occur. The initial random distribution of sectors is rigidly frozen in, and a configuration that nucleates within the habitable island remains there.

The semiclassical transition rate per unit 4-volume, $\gamma(t)$, takes the
standard instanton-mediated form \cite{Coleman1977, ColemanDeLuccia1980, Dowker2002}:

\begin{equation}
\gamma(t) = \frac{\Gamma}{V_4} \sim A\, e^{-S_B/\hbar},
\end{equation}
where $S_B$ is the effective Euclidean action barrier between distinct
topological configurations and $A \sim \mathcal{O}(1)$. To evaluate the
expected number of transitions within a single causally connected patch, we
multiply $\gamma(t)$ by the characteristic spatial 3-volume of a Hubble sphere
($V_3 \sim H^{-3}$), yielding an event rate per unit proper time
\begin{equation}
\frac{dN}{dt} \sim \gamma\, H^{-3}.
\end{equation}
Using the relation $dt = da/(a H)$, where, in keeping with canonical minisuperspace conventions, the scale factor $a$ is defined dimensionally as the physical radius of curvature of the macroscopic spatial slice, the total expected number of topological
transitions occurring after the universe nucleates at an initial scale factor
$a_1$ is
\begin{equation}
N_{\rm tot} = \int_{a_1}^{\infty} \frac{\gamma}{H^4}\, \frac{da}{a}.
\end{equation}

However, the instanton action barrier scales directly with the
characteristic physical volume of the spatial geometry, implying
$S_B \propto H^{-3}$ \cite{ColemanDeLuccia1980, Dowker2002}.
While $\gamma \sim \mathcal{O}(1)$ at the absolute
boundaries of the Planck scale where $H \sim 1$, the rapid drop of the Hubble
parameter during subsequent expansion forces the action barrier to grow
combinatorially ($S_B \to \infty$). Consequently, the factor $\gamma / H^4$
is instantly suppressed, ensuring that $N_{\rm tot} \ll 1$. Virtually no
topological transitions occur post-emergence, rigidly freezing the initial
random distribution of choices into autonomous, non-interfering U-sectors.
We note that these estimates extrapolate semiclassical instanton methods,
developed for bubble nucleation on fixed backgrounds, into the near-Planckian
regime, and should be read as heuristic. The conclusion, however, requires
only the qualitative feature that the action barrier grows monotonically and
rapidly with the physical volume of the transitioning region, which is common
to every known nonperturbative estimate. A dynamical suppression of
topological transitions of this general kind has been argued to arise in
Euclidean quantum gravity on independent grounds \cite{Barvinsky2012}.
The suppression is reinforced kinematically: within a Lorentzian history, a
change of spatial topology forces either closed timelike curves or a
degenerate metric~\cite{Geroch1967}, and trousers-type transitions
concentrate unbounded energy at the join~\cite{AndersonDeWitt1986}.
Topology is therefore chosen at nucleation, where no pre-existing geometry
constrains the choice, and only there; within a history it is not merely
frozen but forbidden.

\subsection{QM Selection of Flux Vectors}

The division of the fundamental spacetime dimensionality ($D$) into macroscopic dimensions ($d$) and microscopic dimensions ($n$) has traditionally been modeled as a dynamical competition between scalar fields (radions) and stabilizing fluxes. We treat the discrete flux vector $\vec N$ as part of the quantum configuration on the same footing as geometry.

Thus, we do not need to fine-tune a radion potential to produce exactly three large dimensions. Instead, the number of macroscopic dimensions is a discrete random variable distributed across the landscape. The crucial difference is that fluxes are \emph{quantized} global data: different $\vec N$ label distinct sectors of the full state. Accordingly, the primordial wavefunction of a manifold capable of magnetic flux is a superposition over flux sectors,

\begin{equation}
	|\Psi_\mathcal{M}\rangle=\sum_{\vec N} c_{\vec N}\,|g_\mathcal{M},\vec N\rangle .
\end{equation}

The same kinetic logic for manifolds applies to flux data. While the relevant suppression scale is set by compact-cycle volumes rather than the macroscopic FRW volume, this is sufficient for rapid freeze-out. Flux-changing processes are nonperturbative and are therefore exponentially sensitive to the characteristic volumes of the cycles they modify: schematically $\Gamma_{\Delta N}\sim \exp[-c\,V_{\rm cyc}]$ (up to prefactors)  \cite{BrownTeitelboim1987}.

In our emergence picture, the compact sector quickly grows from near-Planckian scales into a phase-stable semiclassical configuration, so $V_{\rm cyc}$ increases by many orders of magnitude in fundamental units, driving a correspondingly rapid shutdown of $\Gamma_{\Delta N}$ relative to $H^4$. A sufficiently large flux can backreact on the compact geometry and inhibit this growth, preventing the relevant cycle volumes $V_{\rm cyc}$ from increasing beyond a few Planck lengths.

This suggests a natural cutoff: for each flux choice, there is an effective $N_{\max}$ above which the compact sector cannot enter a large volume, so flux-changing transitions are not quenched. Moreover, because both the energy splitting between neighboring flux sectors and the available decay channels generally grow with $N$, the corresponding transition rates are expected to increase with $N$. As a result, when $N > N_{max}$, the dynamics preferentially drives the system toward smaller flux quanta, until a regime is reached in which $V_{\rm cyc}$ can grow and $\Gamma_{\Delta N}/H^4$ becomes exponentially small.

Finally, it is the mutual backreaction between the emergent manifold geometry and the discrete flux data that drives decoherence into distinct sectors of the theory space,
$\mathcal{T}(\mathcal{M},\vec N)\, $.

Different $\vec N$ generally imply different stress-energy, stabilization conditions, and effective vacuum energies, and hence slightly different semiclassical geometries and expansion histories even when the coarse manifold class is the same. Early on, these macroscopically distinct $(\mathcal{M},\vec N)$ branches entangle with the many inaccessible degrees of freedom of the gravitational field and the compact sector, suppressing interference between them. In this way, the manifold and its flux vectors jointly define the upper hierarchy of branch labels that constitute the ``U-sectors" of the enlarged theory space.

In quantum cosmology, however, one should not think of flux energy as being ``added" to a fixed background. The Wheeler--DeWitt constraint correlates matter/flux content and geometry through the vanishing of the total Hamiltonian,

\begin{equation}
	\big(\hat H_{\rm grav}+\hat H_{\rm flux}\big)\,|\Psi\rangle \approx 0.
\end{equation}

Thus, a branch with larger flux quanta $\vec N$ generally corresponds to a slightly different semiclassical geometry (and, in unimodular language, a different compensating value of $\Lambda$ as an integration constant) so that the total constraint is satisfied for the \emph{same} coarse manifold class. In this sense the ``energy" associated with flux is balanced by the gravitational sector through the constraint itself.

\subsection{Matter Fields as Internal Geometry}

Allowing extra dimensions natively enlarges the geometric data beyond the macroscopic metric. In this framework, the gauge fields and matter representations observed in the four-dimensional effective theory are not externally added components; they are strictly the geometric properties of the internal compactification manifold. Just as the macroscopic metric dictates classical gravity, the isometries of the extra dimensions manifest as gauge forces, and the zero-modes of the internal Dirac operator manifest as chiral matter fields.

To specify a candidate semiclassical background, the wavefunctional must account for the specific state of this internal geometry. This state is parameterized by connections on internal principal bundles ($\mathcal{B}$) and continuous geometric proportions ($\lambda$); both housed within the topological container $(\mathcal{M})$.
\begin{itemize}
    \item The bundle connections ($\mathcal{B}$) dictate the exact gauge symmetries and matter representations, determining the available ``species'' of particles and forces.
    \item The continuous moduli ($\lambda$) determine the cycle volumes and manifold shapes, which dynamically fix the physical constants, gauge couplings, and mass hierarchies.
\end{itemize}

The fundamental superposition remains defined entirely by the generative basis:

\begin{equation}
|\Psi\rangle = \sum_{\mathcal{M}, \vec{N}} c_{\mathcal{M}, \vec{N}} \, |\mathcal{M}, \vec{N}\rangle
\end{equation}

As the universe enters a semiclassical expanding regime, the specific bundle structures ($\mathcal{B}$) and geometric proportions ($\lambda$) assumed within each $(\mathcal{M}, \vec{N})$ sector dictate vastly different effective actions. These macroscopic differences in expansion histories and localized field dynamics rapidly entangle with the environment, driving the decoherence of the superposition into distinct, non-interfering U-sectors.

\subsection{Moduli}

The presence of the quantized flux $\vec{N}$ introduces a potential energy term $V_{flux}$ that depends on the geometry of the extra dimensions (moduli fields). Physically, this represents the energy cost of compressing the magnetic field lines, acting as a repulsive potential that prevents the extra dimensions from collapsing~\cite{GKP2002, DouglasKachru2007}:

\begin{equation}
	V_{flux}(R) \sim \frac{|\vec{N}|^2}{R^{2(p+1)}}.
\end{equation}
where $R$ is the characteristic radius of the flux-threaded cycle (so that its volume is $V_{cyc} \sim R^p$) and $p$ is the dimension of that cycle. The precise exponent depends on $p$, but the qualitative behavior, a repulsive wall diverging as $R \to 0$, holds for any $p \geq 0$ and is what stabilizes the cycle against collapse.

Simultaneously, gravity provides an attractive tension.  As the spatial geometry initiates its semiclassical expansion, the moduli are immediately subjected to the combined effective potential $V_{eff}(\chi)$ established by the discrete flux vectors $\vec{N}$, causing them to dynamically ``roll" toward their minima. Moduli are continuous scalar fields that are determined by the geometric and topological data of the compact dimensions (sizes of handles and loops, shapes, complex structure, overall size, \emph{etc.}). When stabilized, their vacuum expectation values $\langle \chi^A \rangle$ , where the index A runs over the dimensions of the moduli space, determine the strengths of interactions and the blueprint for mass generation.

Dynamics of Stabilization: It is crucial to distinguish the stabilization of these continuous fields from the discrete freezing of the topology. While the flux vector $\vec{N}$ is rigidly locked at the moment that the manifold is established ($a \sim$ several $\ell_P)$ due to the volumetric suppression of tunneling rates, the moduli fields $\chi^A$ remain dynamical in the immediate post-Planckian epoch. They effectively ``roll" within the potential $V_{eff}(\chi)$ created by the fluxes, oscillating until the Hubble expansion rate drops below their effective mass $(H < m_\chi)$.

The shape and depth of the moduli potentials depend in a complicated way on the hundreds of flux vectors; consequently, there are an enormous number of possible locations where the potential minima can be and hence an effective continuum of values for particle masses and force laws~\cite{BoussoPolchinski2000}. The stabilization of geometric moduli proceeds hierarchically. In standard flux compactifications, the discrete flux vectors $\vec{N}$ induce a classical Gukov-Vafa-Witten superpotential. Rather than acting as hard geometric boundaries, this superpotential generates a smooth, steep scalar potential that stabilizes the complex structure moduli and the dilaton. These fields acquire heavy masses dynamically bounded by the Kaluza-Klein scale, ensuring they rapidly settle into their minima and decouple from the low-energy effective field theory \cite{DouglasKachru2007, Denef2008}.

If we assume radiation-dominated expansion $t \propto a^{1/2}$, then $H \propto 1/a^2$, and these moduli settle at $a < 100 \ell_{\rm Pl}$, dissipate their initial quantum fluctuations, and settle into the vacuum minimum $\chi_*$ prior to the onset of cosmic inflation. However, moduli with lighter $m_\chi$ that do not fall to the potential minimum before inflation (i.e., if $ m_\chi < H_{\text{inflation}}$), will be ``Hubble locked" by cosmic friction at a random displacement away from the minimum $\chi_*$.

Eventually, when H drops below the mass of the modulus ($H < m$), the friction vanishes, and it begins to oscillate in sync everywhere in the universe. This misalignment mechanism creates a secondary source of diversity: while the discrete flux vector $\vec{N}$ determines the shape of the physical laws (the potential landscape), the specific values of lighter parameters (such as axionic angles or soft masses) may be determined by the stochastic initial conditions frozen in by inflation\cite{Kachru2003}.

\section{The Hyperdimensional Feynman Sum of Geometries}
\label{sec:hyperdimensional}

To fully capture the generative capacity of the Wheeler--DeWitt framework, the functional integral cannot be artificially restricted to a fixed four-dimensional Einstein-Hilbert background. Kinematically, the universal wavefunction $\Psi$ is a superposition over the
entire space of geometric configurations, every dimension and every topology,
with no dynamics presupposed. This kinematic Hilbert space is the
unconstrained parent of the theory. The physical content of $\Psi$ is the
subset that satisfies the Wheeler--DeWitt constraint, and the selection is not
performed by a Hamiltonian standing prior to the geometries. Rather, each
geometric configuration induces its own constraint operator $\hat{H}_{\mathcal{M}}$,
a functional of that geometry's dimension, supermetric, and curvature. A
configuration is physical only if it is self-consistent: only if it is
compatible with the dynamics it itself generates. The viable U-sectors are
exactly these self-consistent geometries.

The Everettian multiplicity of universes must be formalized through a generalized master partition function. This approach sums over all possible topologies, dimensionalities ($D$), and geometric shapes, treating the generation of distinct physical realities as a quantum-mechanical superposition.

Because the Wheeler--DeWitt constraint ($\hat{H}\Psi = 0$) dictates a timeless state, the universe does not nucleate and dynamically evolve ``forward" relative to an external time parameter. The quantum generation of these U-sectors is properly formulated as a Feynman sum over geometries evaluated from the absolute initial boundary $\mathcal{B}_Q$. The transition amplitude is given by the hyperdimensional sum.

\begin{equation}
	\begin{split}
	\Psi[h_{ij}, \Phi] =  \mathcal{N}_{\mathrm{grand}} \sum_{D=D_{\mathrm{min}}}^{\infty}  \mathcal{N}_D \sum_{\mathcal{M}\in\mathcal{T}_D} \mathcal{N}_{\mathcal{M}} \\ \int_{\mathcal{B}_Q}^{h_{ij}} \mathcal{D}(g_D) \, e^{\frac{i}{\hbar} S_{\mathrm{geom}}[g_D]} \, \mathcal{Z}_{\mathrm{matter}}[g_D, \Phi]
	\end{split}
\label{eq:sumofgeometries}
\end{equation}
where $h_{ij}$ is the induced 3-metric on the spatial slice and $\Psi[h_{ij}]$ is a functional on superspace. The matter partition function is nested within the geometric integral, strictly dependent on the underlying metric background:
\begin{equation}
\mathcal{Z}_{\mathrm{matter}}[g_D, \Phi] = \int^{\Phi} \mathcal{D}(\phi_D) \, e^{\frac{i}{\hbar} S_{\mathrm{matter}}[g_D, \phi_D]}
\end{equation}

$\Psi[h_{ij}]$ assigns a complex probability amplitude to a specific intrinsic 3-geometry. Its squared norm, $|\Psi[h_{ij}]|^2$, defines the relative statistical weight of the U-sector containing that spatial slice within the global Many-Worlds superposition.
In this formulation, a strict conceptual distinction must be made between the measure, the nested normalizations, and the action. The Normalization Hierarchy: The total universal state requires a grand normalization factor ($\mathcal{N}_{\mathrm{grand}}$). Because the path integral evaluates an orthogonal superposition of fundamentally distinct physical laws, the amplitude must be normalized layer by layer. Each dimensionality introduces a dimensional weight ($\mathcal{N}_D$), and within each dimension, each distinct topological manifold requires a specific sector normalization ($\mathcal{N}_{\mathcal{M}}$). These hierarchical coefficients define the relative statistical weight of each macroscopic branch in the landscape.

The Matter Partition Function ($\mathcal{Z}_{\mathrm{matter}}$): The matter measure $\mathcal{D}(\phi_D)$ evaluates the quantum fluctuations of all fields and discrete fluxes strictly upon the specific macroscopic geometry $g_D$ being integrated. This formalizes the dependence of the Standard Model on the spacetime topology. Crucially, if the continuous geometric manifold fails (as at the $\mathcal{B}_Q$ boundary, where QTH proliferation destroys extended differentiable Lorentzian connectivity), the background required to evaluate $\mathcal{Z}_{\mathrm{matter}}$ ceases to exist. The effective field theory on that continuing macroscopic spacetime terminates; this does not require the disappearance of every local admissible island.
What survives is therefore not selected by an external rule but by
self-consistency: a sector is real precisely when its geometry and the
constraint that geometry induces close upon a common, normalizable state.

The Action as a Structural Filter: The exponential action ($e^{iS/\hbar}$) acts purely as the structural filter over the unconstrained generative space. It is this filter that prevents the integral from defaulting to infinite dimensions. Consequently, macroscopic hyperdimensional spaces are natively suppressed, allowing only a discrete spectrum of finite-dimensional, topologically distinct, and stable block manifolds to emerge. What survives this hierarchical filter is a coherent quantum superposition of disjoint U-sectors, each governed by its own effective field theory.

This transition amplitude formulation radically resolves the paradox of the cosmic initial boundary. Because the path integral generates a complete, static block universe, time is strictly an internal, emergent property of the generated manifold $(h_{ij})$. The $\mathcal{B}_Q$ boundary does not represent a chronological moment preceding the universe. Rather, it is simply the absolute lower limit of the functional integration. It represents the structural threshold, emerging at the boundary of the late Planck era, where expanding geometric cells percolate into a smooth space. Because the Wheeler--DeWitt equation rigorously prevents the formation of pre-geometric spacetime foam at this limit, it establishes a definitive origin for a standard big bang. The traditional question of ``what came before" the universe is mathematically preempted, as the entire temporal foliation of a given U-sector is strictly defined and contained entirely within the limits of the integral.

\subsection{Dimensional Selection and Anthropic Viability}
\label{sec:dimensional}

The inclusion of the sum over dimensionality ($\sum_D$) formally implements a kinematic sector decomposition. Distinct dimensionalities induce inequivalent constraint operators and, with them, inequivalent field content and operator algebras; the operator is fixed by the geometry, not imposed upon it. There is no single operator algebra acting across different $D$-sectors.  One may express this as the vanishing of cross-dimensional matrix elements for any local observable $\hat{\mathcal{O}}$:

\begin{equation}
	\langle \Psi_{D'} \,|\, \hat{\mathcal{O}} \,|\, \Psi_{D} \rangle = 0 \qquad \text{for}\qquad D'\neq D,
	\label{eq:D_superselect}
\end{equation}

While the functional integral evaluates sectors with all possible values of $d \leq D$, the vast majority of these branches are structurally sterile regarding complexity. Following Ehrenfest's classical arguments, branches with $d > 3$ yield unstable orbits ($V(r) \propto r^{-(d-2)}$) that forbid stable bound states, while $d < 3$ provides insufficient topological complexity for neural connectivity and violates Huygens' principle for clean wave propagation. Therefore, while the hyperdimensional sum populates a vast landscape, the only anthropically viable branches are those with macroscopic dimensions of $d=3$.

In this Feynman propagation integral, relationships can be found between the measure and the action. The non-perturbative measure, $\mathcal{D}(g_D) \mathcal{D}(\phi_D)$, supplies the full, unconstrained generative space, iterating over all valid dimensionalities, discrete topologies, and matter configurations. The action, $e^{iS/\hbar}$, acts purely as the structural filter. It is this filter that prevents the integral from defaulting to infinite dimensions. Higher-dimensional manifolds naturally possess much larger actions ($S_D \gg \hbar$), which subjects them to violent phase cancellation in the path integral. Consequently, macroscopic hyperdimensional spaces are natively suppressed, allowing only a discrete spectrum of finite-dimensional, topologically distinct, and stable block manifolds to emerge.

\subsection{The Underlying Geometric Universe}

If we accept that matter fields are not fundamental entities, but rather the geometric projections of higher-dimensional topologies into our 4D spacetime, then the standard Hamiltonians cluttered with disparate particle and field terms are merely phenomenological approximations. In this view, the complexity of the Standard Model (all those gauge bosons and fermion species) is simply a mathematical transformation, useful for low-energy calculation but physically redundant. The true master Hamiltonian is a purely geometric construct, acting on the totality of the manifold’s configuration space.

Rather than postulating an external matter-inclusive Lagrangian, we define the theory via the constrained canonical algebra of the vielbein-connection pair $(e, \omega)$. In this framework, the Hamiltonian is not an energy-provider, but the \emph{universal generator of gauge transformations}, the mathematical guardian that ensures physical law is invariant under any choice of coordinate slicing. The Einstein-Cartan Hamiltonian, encompassing both curvature and torsion-induced spin-density, serves as the complete engine for the Wheeler-DeWitt equation across all sectors of the Universal Wavefunction $\Psi$.

The topology of the manifold dictates the Hamiltonian's structure: it is the operator that measures the energy of a connection-based geometry. We know from the Einstein-Cartan identity that the curvature scalar $R(\omega)$ automatically decomposes into the Ricci scalar and the quadratic torsion terms:
\begin{equation}
R(\omega) = R(\text{Levi-Civita}) + \alpha |T|^2
\end{equation}
Because the manifold's topology mandates the existence of a spin structure in these sectors, the Hamiltonian naturally manifests the spinor energy term as the torsion-based response required for closure. Schematically (suppressing the index structure, densitization, and decomposition into Hamiltonian and momentum constraints), the constraint takes the form:
\begin{equation}
\hat{\mathcal{H}} = \frac{\kappa}{2} \pi_{ab}^{\mu\nu} \pi_{\mu\nu}^{ab} - \frac{1}{\kappa} \sqrt{g} R(\text{LC}) + \bar{\psi} \left( -i \gamma^i \nabla_i + m \right) \psi = 0
\end{equation}

Under this paradigm, the entirety of the multiverse is captured by a path integral over all admissible geometric configurations. The resulting action acts as the fundamental arbiter of physical law:
\begin{equation}
S = \int d^D X \sqrt{-g} \left[ \frac{1}{2\kappa} R(\omega) + \alpha |T|^2 \right]
\end{equation}
This single, purely geometric action generates the existence of all distinct sectors within the multiverse. Through the Feynman sum over geometries, the constructive interference of these geometric configurations establishes the specific physical laws and topological structures inherent to each sector. In this framework, ``laws of physics" are not prescriptive rules added to the manifold, but rather the emergent, static properties of the geometry itself.

The quantum boundary restricts the gravitational sum to histories supporting an extended differentiable Lorentzian geometry. The relevant restriction concerns the full finite-scale metric response to curvature and quantum fluctuations. In spatially flat radiation FRW geometry, $R=0$ and the Weyl tensor vanishes, but $S_{\hat\mu\hat\nu}=R_{\hat\mu\hat\nu}=(H^2/c^2)\operatorname{diag}(3,1,1,1)$ and $R_{\mu\nu}R^{\mu\nu}=12H^4/c^4$ are nonzero. Thus vanishing scalar curvature does not remove the homogeneous curvature response that, together with fluctuations, drives elementary regions outside Lorentzian signature. In Ricci-flat regions, Weyl curvature supplies a distinct tidal response. In either case, QTH proliferation can eliminate extended differentiable Lorentzian connectivity and terminate the history at $\mathcal{B}_Q$. This provides a geometric admissibility restriction on the configurations entering the sum; convergence and renormalizability of the resulting functional integral require separate analysis.

In the purely geometric description proposed here, a fermion is represented by a \emph{spinor knot}, a localized, topologically nontrivial configuration of the torsion sector. Such a configuration must have a finite differentiable core supported by admissible metric elements; its continuation cannot be assumed through a region where QTH proliferation eliminates the required Lorentzian support.

\subsection{Gravitational-Wave Detectability}
\label{subsec:gw-detectability}

A possible observational consequence concerns the initial graviton population. For this purpose alone, suppose that the two propagating four-dimensional graviton polarizations have a thermal distribution at nucleation, with physical temperature $T_{\rm qb}$. This is an additional assumption about the tensor state; the vacuum fluctuations used in the signature calculation do not by themselves establish a thermal population.

Let $f_g$ denote the fraction of the boundary energy density carried by these thermal gravitons, so that $\rho_g=f_g\rho_{\rm qb}$. In the relativistic equilibrium-gas approximation,
\begin{equation}
 \rho_g c^2=\frac{\pi^2}{15}
 \frac{(k_B T_{\rm qb})^4}{\hbar^3c^3},
\end{equation}
and therefore
\begin{equation}
 \frac{T_{\rm qb}}{T_P}
 =\left[\frac{15 f_g}{\pi^2}
 \frac{\rho_{\rm qb}}{\rho_P}\right]^{1/4}.
 \label{eq:Tqb-gravitons}
\end{equation}
For $q=2.5\ell_P$ and $C_{\rm FRW}=1$, the measured mean boundary density gives $T_{\rm qb}\simeq0.38 f_g^{1/4}T_P$. The choice $f_g=1$ is a graviton-dominated benchmark. No contribution from particle species that have not yet emerged is included in this estimate.

For a thermal tensor initial state, the result of Bhattacharya, Mohanty, and Nautiyal \cite{BMN2006} gives, in units $\hbar=c=k_B=1$,
\begin{equation}
 P_T(k)=P_T^{\rm vac}(k)
 \coth\!\left(\frac{k}{2\Theta_g}\right),
 \qquad \Theta_g\equiv a_{\rm qb}T_{\rm qb},
 \label{eq:coth}
\end{equation}
where $k$ is the comoving wavenumber and $a_{\rm qb}$ is the scale factor at nucleation in the same comoving convention. The quantity $\Theta_g$ is the redshifted comoving graviton temperature for freely propagating massless modes. The associated comoving crossover scale is
\begin{equation}
 k_T=2a_{\rm qb}T_{\rm qb}.
 \label{eq:kT}
\end{equation}
For $k\gg k_T$, the vacuum spectrum is recovered; for $k\ll k_T$, the multiplier approaches $2\Theta_g/k$. The observable would be a scale-dependent enhancement of the primordial tensor spectrum, potentially appearing in large-angle CMB $B$ modes.

Whether this scale lies within the observable window depends on $f_g$, the boundary density, and the expansion between nucleation and observation, including the duration of inflation and reheating history. Additional inflation redshifts the characteristic scale to longer present wavelengths. No fixed cutoff in the number of extra e-folds follows from the boundary-density calculation alone. A measured enhancement consistent with this profile could constrain the initial graviton state and its redshifted temperature; it would not uniquely establish this nucleation mechanism. Absence of an enhancement could reflect a nonthermal tensor state, a smaller graviton energy fraction, or a crossover outside the accessible range.

\subsection{The Universal Crunch}
\label{sec:Cyclic}

It is necessary to reframe the mechanics of a universal crunch in light of the $\mathcal{B}_Q$ boundary \cite{DeWitt1967, Shaya2026_Sig}.
In a late-stage contracting universe, the manifold is either already densely fragmented by astrophysical black holes and their topological boundaries or heated by contraction into a homogeneous fluid. In the homogeneous radiation limit, the trace-free Ricci response combines with quantum metric fluctuations to generate QTHs and destroy differentiable Lorentzian connectivity. Collapse ends at a late-time quantum boundary $\mathcal{B}_Q$ if no new dynamics reverses the contraction before that threshold. The next subsection compares the illustrative boundary densities with a standard bounce scale.

Because an astrophysical black hole forms a topological void in this framework, rather than a microscopic singularity, the terminal collapse of a fragmented universe is not the violent compression of a smooth space into a global singularity, nor is it driven by the gravitational attraction of point masses. In the late phases of collapse, all of the event horizons merge, swallowing all of space into a unified trapped region. Because the center of each horizon had been a quantum boundary, a topological void, these voids exert no direct active kinematic pull. Instead, the dynamics are governed entirely by the residual smooth trapped metric field between them.

At the terminal boundary, fluctuations combined with curvature-induced deformation have taken a substantial fraction of elementary regions outside the Lorentzian class. Their exclusion as QTHs removes the extended differentiable connectivity needed for a continuing spacetime and temporal foliation. This endpoint is not attributed to the action reaching $\hbar$ or to dissolution of a saddle. The Wheeler--DeWitt constraint ($\hat H\Psi=0$) provides the timeless configuration-space description used here; Schr\"odinger evolution with an emergent time applies within the surviving semiclassical spacetime domain. The block universe of the sector ends at loss of admissible Lorentzian continuation, the same criterion that defines its origin in the time-reversed interpretation (Sec.~\ref{subsec:WDWDomain}). The unitary normalization of the universal wavefunction is maintained not by serial continuation through a singular point, but by the parallel, timeless existence of the vast ensemble of disjoint U-sectors within the extended configuration space.  The probability amplitude is not distributed over space; it is distributed over all possible 3-geometries ($h_{ij}$) and matter configurations ($\phi$). The normalization condition is an integral over the infinite-dimensional configuration space (Superspace):
\begin{equation}
	\int |\Psi[h_{ij}, \phi]|^2 \mathcal{D}[h] \mathcal{D}[\phi] = 1
\end{equation}
Notice that time does not appear in this integral.  Some U-sectors will survive for finite times.

Disconnected Lorentzian fragments surviving the loss of the parent universe's extended connectivity can seed baby universes through renewed growth. If reduced to one or a few elementary regions, these fragments would retain little or no structural information about the parent universe.

\subsection{The Preemption of Bounce Cosmologies}
\label{subsec:bounce}

A natural question is whether a contracting universe can avoid terminal truncation by undergoing a smooth bounce. Within the present framework, the relevant comparison is between the density at which new dynamics reverses contraction and the density at which the adopted response model loses extended differentiable Lorentzian support.

For reference, the standard improved-dynamics effective model of loop quantum cosmology (LQC) has critical density
\begin{equation}
  \rho_c = \frac{\sqrt{3}}{32\pi^2\gamma^3}\,\rho_P \simeq 0.41\,\rho_P,
  \label{eq:rhoc}
\end{equation}
for the Barbero--Immirzi parameter $\gamma\simeq0.2375$ \cite{Ashtekar2006}. By comparison, the homogeneous radiation scans described in Sec.~\ref{subsec:NativeEmergence}, with $C_{\rm FRW}=1$, give
\begin{equation}
  \frac{\rho_{\rm qb}}{\rho_P}\simeq
  \begin{cases}
    0.01982, & q=1.9\ell_P,\\
    0.01317, & q=2.5\ell_P,\\
    0.005988, & q=4\ell_P.
  \end{cases}
  \label{eq:rhoBQ}
\end{equation}
These illustrative boundary densities lie approximately 21 to 68 times below the quoted LQC critical density. If the adopted Lorentzian admissibility and finite-scale response remain applicable during contraction, and bounce modifications have not already changed that response or reversed the contraction, the geometry loses extended differentiable Lorentzian support before reaching the standard LQC bounce density.

This is a conditional preemption argument, rather than a universal exclusion of bounce cosmologies. A mechanism that reverses contraction below the relevant signature-connectivity threshold could avoid the boundary. Likewise, new dynamics that changes the quantum metric ensemble or its curvature response requires a new connectivity calculation. Once extended differentiable Lorentzian support is lost, however, a smooth continuation through a bounce is unavailable within the class of histories admitted here. No assertion that the action falls to $\hbar$ or that a stationary-phase saddle dissolves is needed for this signature-based termination criterion.

\section{The Measure Problem}
\label{sec:measures}

The relative amplitudes of admissible nucleating geometries remain to be determined. In particular, the framework does not establish a preference for homogeneous initial conditions or low Weyl curvature. The homogeneous radiation geometry used in the illustrative quantum-boundary calculation is an assumed background, rather than an outcome of a nucleation-weight calculation.

The framework implies a cosmology of vast cardinality, unifying the landscape
of string theory with the branching of Everett: the flux-compactification
landscape alone implies of order $10^{500}$ sectors. A single-universe model
must fine-tune to land in a habitable one; here, the universal wavefunction
$\Psi$ carries non-zero amplitude across the whole ensemble. The anthropic
content of this picture is correspondingly modest. It requires only that $\Psi$
assign \emph{non-zero support} to the habitable region of theory space.
Conditioning the probability on the presence of an observer renders our specific location unsurprising.
No normalized measure over the
landscape is needed for this step; however minute, and even if not sharply
defined, the habitable island need only be realized with non-zero amplitude.

It would be too quick to conclude that landscape measures are therefore without
use. What the existence argument renders irrelevant is the \emph{absolute}
measure of the habitable island, its size as a fraction of the whole. The
\emph{relative} measure across classes of manifolds is another matter, and it
serves as an essential guide for self-location.  If life-supporting islands concentrate in
manifolds of a particular type, anthropic conditioning naturally localizes our expected position to those dominant manifolds, the same self-locating logic that underlies Weinberg's anthropic estimate of the cosmological constant~\cite{Weinberg1987}. 
Concrete questions of this kind are well-posed for mapping the topography of the landscape.

Do we live in a Calabi--Yau compactification, a $G_2$ manifold, or some other
class, and does one of these carry systematically greater life-measure? What
mechanisms generate a dark sector, and do they raise the complexity measure of
the manifolds that host them? And, as a direct consistency test of the whole
construction, are there life-forms governed by greatly different laws that inhabit a vastly larger phase space than ours? Were the measure found to place our own type of physics many orders of magnitude more rare than other habitable sectors, it would not strictly falsify the framework, as the universal wavefunction guarantees the existence of statistical outliers. However, such a result would impose a severe Bayesian penalty, requiring an extraordinarily strong prior to justify our anomalous position. The measure problem is thus worth pursuing not to falsify the landscape, but to accurately map its topology and locate our specific thermodynamic sector within it.

To make the term habitable precise without privileging our own chemistry, we adopt
a substrate-independent criterion: life requires that stored non-trivial information persist
far longer than the environmental dynamical timescale (e.g., kinematic scattering, foldings, reactivity),
\begin{equation}
  \tau_{\rm mem} \;\gg\; \tau_{\rm dyn}.
  \label{eq:life}
\end{equation}
This is a necessary condition, not a sufficient one, but it suffices to
partition theory space, and it is computable from a sector's effective theory
without any assumption about what its life forms resemble. The dimensional
selection of Subsection~\ref{sec:dimensional} is simply
Eq.~\eqref{eq:life} applied at the scale of orbital dynamics: only in $d = 3$ spatial dimensions do stable bound configurations exist in which information
can be stored at all, so every other dimensionality fails Eq.~\eqref{eq:life}
at the most basic level, with $\tau_{\rm mem}$ collapsing to the dynamical time.
The unique selection of $d=3$ as the sole dimensionality permitting stable, bound configurations is the simplest, most immediate confirmation of this predictive logic.

For a \emph{single} sector with specific manifold $(\mathcal{M}_n, \vec{N})$, the probability that it supports complexity
is well defined and computable in principle: the fraction of the sector's
moduli space on which Eq.~\eqref{eq:life} holds, evaluated against the natural
measure the moduli space inherits from the kinetic terms of the effective
action. This per-sector quantity, $p(\mathcal{M}_n, \vec{N})$, is the largest
object in the problem that is cleanly defined; its difficulties are the ordinary
ones of measure theory on a finite-dimensional space. The \emph{aggregate} over
all sectors is another matter. The number of admissible $(\mathcal{M}_n,
\vec{N})$ is enormous and possibly unbounded; the count grows faster than any
fixed suppression, and no closed-form measure over the set of topologies is
known. The absolute, landscape-wide probability of life is therefore not
presently computable, and may not be well defined at all.

We do not claim to resolve this aggregation.  The gap between the well-defined per-sector
probability and the undefined aggregate is an
empirical frontier. Each sector studied yields a datum $p(\mathcal{M}_n,
\vec{N})$, and the accumulation of such data, a direct extension of the
statistical study of the string landscape~\cite{Douglas2003, AshokDouglas2004},
may, if regularities exist, reveal how complexity-probability depends on
topological invariants, and so enable an inference about the ensemble, and the
relative measures that direct summation cannot provide. 
Whether such regularities exist is an open question.

\section{Robustness Against Objective Reduction Mechanisms}
\label{sec:OR_baseline}

If a gravitational collapse mechanism, such as Penrose's Objective Reduction~\cite{Diosi1989,Penrose1996}, were to be confirmed in some regime, it would likely be compatible with the Topological MWI framework proposed here. The Penrose criterion is formulated for superpositions of alternative mass distributions within a given semiclassical spacetime, quantified by a gravitational self-energy difference $\Delta E_G$.

In this framework, where the superposition involves distinct global topologies ($\mathcal{M}$) rather than local mass displacements, the classical definition of the gravitational self-energy difference breaks down. Consequently, we argue that the collapse mechanism would not act on the universe's wavefunction during the generative phase. The topological branches would essentially be energy-degenerate vacuum states, allowing the superposition to persist and evolve unitarily until the geometric degrees of freedom freeze out, well before any effective gravitational collapse mechanism could intervene.

Therefore, the generation of the U-sector landscape and its subsequent resolution of the fine-tuning problem remain robust even if the universal validity of the Many-Worlds Interpretation is relaxed in favor of objective reduction in later, local environments.

\section{Conclusions}
\label{sec:Conclusions}

The results of this paper follow from a single premise -- the universal
validity of unitary quantum mechanics -- applied to a Wheeler--DeWitt path
integral that sums over topologies. From that premise, the topological
multiverse is not imposed but implicit. Its notable accomplishments are the
following.

\begin{itemize}

\item \textbf{A multiverse that is derived, not postulated.} Everettian quantum
mechanics, applied to a path integral that admits topology change, already
produces a multiverse of disjoint U-sectors. It requires only three ingredients
the formalism already supplies: unitary evolution, a sum over admissible
geometries, and the superselection of topological labels, from which the
direct-sum state $\Psi = \sum_{\mathcal{M},\vec{N}} c_{\mathcal{M},\vec{N}}
|\mathcal{M},\vec{N}\rangle$ follows as a self-consistent reading rather than an
added postulate. Single-universe cosmology is then the position that must
\emph{add} structure to suppress what the formalism already generates.

\item \textbf{Constants of nature promoted to dynamical variables.} Enlarging
the DeWitt sum to include all smooth manifolds supporting causal dynamics turns
the dimensionality, gauge groups, and couplings from fixed background inputs
into quantum variables of the sum, each U-sector carrying a complete set of
effective laws and its own block of the constraint operator.

\item \textbf{A non-singular origin at $\mathcal{B}_Q$.} In the final moments of a collapsing universe, quantum metric fluctuations combined with curvature-induced deformation drive a substantial fraction of elementary metric regions outside the nondegenerate Lorentzian class. Their exclusion as QTHs destroys the extended differentiable Lorentzian connectivity of the parent universe. The boundary is set by loss of Lorentzian metric signature and connectivity, rather than by reaching a singularity. Disconnected Lorentzian remnants may survive and seed baby universes, as discussed in Sec.~\ref{sec:Cyclic}.

In the time-reversed expanding interpretation, it marks the onset of connected admissible spacetime. 

\item \textbf{Dynamical freeze-out of the sectors.} Inter-sector transitions are
suppressed shortly after nucleation, freezing the initial distribution of
topologies into autonomous, non-interfering U-sectors and preventing a settled habitable configuration from tunneling away.

\item \textbf{Conditional preemption of the bounce.} For $C_{\rm FRW}=1$, the homogeneous radiation scans place the quantum boundary at approximately $0.006$ to $0.020\,\rho_P$ for the tested elementary sizes. These densities are below the standard effective LQC bounce density, $0.41\,\rho_P$. Within the adopted response model, the parent universe loses extended Lorentzian connectivity before that bounce scale unless new dynamics changes the response or reverses contraction at a lower density.

\item \textbf{A possible gravitational-wave signature.} Section~\ref{subsec:gw-detectability} relates an assumed graviton initial state to a possible scale-dependent tensor enhancement. Its observability depends on the initial graviton energy fraction and subsequent expansion history.

\item \textbf{A unified, substrate-independent habitability criterion.} Life is
characterized by the persistence of stored information against disruption,
$\tau_{\rm mem} \gg \tau_{\rm dyn}$. The selection of three spatial dimensions
follows as the same criterion at the scale of orbital dynamics, $d=3$ being the
unique dimensionality admitting the stable bound states in which information can
persist.

\item \textbf{A delimited measure program.} The anthropic conclusion requires
only non-zero support on the habitable set, not a global landscape measure. The
probability of complexity on a single sector is well defined and computable;
the aggregate over sectors is open, and is left to a concrete empirical program
in the spirit of landscape statistics~\cite{Douglas2003, AshokDouglas2004},
whose relative measures across manifold classes, if regularities emerge,
could become predictive for our own physics. Relative nucleation amplitudes and the origin of large-scale homogeneity remain open questions.

\end{itemize}

\noindent The phenomenological complexity required for observers is thus traced
to one foundational principle. The path integral provides its calculational
language, and Many-Worlds~\cite{Everett1957} its literal interpretation: the
topological multiverse is what unitary quantum mechanics, applied to a
gravitational sum over geometries, already contains.

\bibliographystyle{apsrev4-2}
\bibliography{mwicosmo}
\end{document}